\documentclass[]{pasj01}
\draft

\begin{document} 
\Received{}%{yyyy/mm/dd}
\Accepted{}%{yyyy/mm/dd}
%\Published{yyyy/mm/dd}

\title{Iron emission line from the spiral galaxy M101}

%%% begin:list of authors
% Do NOT capitalize all letters in "textsc".
\author{Shigeo \textsc{Yamauchi}\altaffilmark{}%
%\thanks{Example: Present Address is xxxxxxxxxx}
}
\altaffiltext{}{Department of Physics, Nara Women's University, Kitauoyanishimachi, Nara 630-8506}
\email{yamauchi@cc.nara-wu.ac.jp}
%%% end:list of authors

%% `\KeyWords{}' always has to be placed before `\maketitle'.
\KeyWords{galaxies: individual (M101) --- galaxies: ISM --- X-rays: galaxies --- X-rays: ISM} %Do NOT move this preamble from here!

\maketitle

\begin{abstract}
Archival Suzaku data of the face-on spiral galaxy M101 were analyzed. 
An intense emission line at 6.72$^{+0.10}_{-0.12}$ keV was detected 
in the central region. 
This line is identified with a K-line from He-like iron, 
which indicates the existence of a thin thermal plasma with a temperature of several keV. 
The iron line luminosity within the central 5 arcmin radius region is estimated to be (2--12)$\times$10$^{37}$ erg s$^{-1}$. 
The origin of the iron emission line is discussed.
\end{abstract}

\section{Introduction}

The study of X-ray properties of spiral galaxies was made possible by the Einstein satellite (e.g., \cite{Fabbiano1989}).
X-ray observations with spatial resolution resolved individual X-ray sources 
and revealed the existence of diffuse X-ray emission with $kT$$<$1 keV
(e.g., \cite{Fabbiano1989,Fabbiano1992,Read1997,Tyler2004,Owen2009,Mineo2012}).
The diffuse X-ray emission is considered to originate from gas shock-heated by supernova (SN) explosions 
and stellar wind interactions and made by star formation activity.
Fe-K lines at $\sim$6.7 keV were detected from 
starburst galaxies (e.g., \cite{Pietsch2001,Boller2003,Ranalli2008,Mitsuishi2011}) and ultraluminous infrared galaxies 
(e.g., Iwasawa et al. 2005, 2009), indicating the existence of a hot plasma with a temperature of $>$1 keV.
The Fe-K line was also detected from the central region of the normal spiral galaxy M31 \citep{Takahashi2004}, 
while the line has not been found in any other normal spiral galaxies so far.

Unresolved thermal X-ray emission with an intense Fe-K line has been found in the Milky Way (MW). 
This is called Galactic Diffuse X-ray Emission (GDXE). 
The GDXE is found in the Galactic disk 
(e.g., \cite{Worrall1982,Warwick1985,Koyama1986,Yamauchi1993,Kaneda1997,Sugizaki2001}), 
the Galactic center (e.g., Koyama et al. 1989, 1996, 2007b; Yamauchi et al. 1990), 
and the Galactic bulge regions (e.g., \cite{Yamauchi1993,Kokubun2001,Revnivtsev2003}).
Although information on the properties of the GDXE has increased since its discovery, its origin remains unsolved.
The most important issue is whether the GDXE is truly diffuse emission or 
composition of numerous faint X-ray sources.
If the diffuse origin is correct, the hot plasma 
having a huge thermal energy is hard to be confined by the Galactic gravity. 
On the other hand, if the point source model is correct, 
a large number of sources having a thin thermal emission 
($kT$=5--10 keV) with intense Fe-K emission lines are required.

The Fe-K emission line may be a common feature among galaxies. 
Thus, research for spiral galaxies is valuable for not only exploring the activities of the galaxies 
but also understanding the GDXE.  

M101 (NGC 5457) is a nearby face-on SAB(rs)cdI galaxy \citep{deVaucouleurs1991} 
at the distance of 6.8 Mpc \citep{Saha2006}. 
It is located at the high Galactic latitude, ($l$, $b$)=(\timeform{102D.0}, \timeform{+59D.8}) and 
the Galactic absorption is low ($N_{\rm H}$=1.16$\times$10$^{20}$ cm$^{-2}$, Dickey \& Lockman 1990).
M101 is a vigorously star-forming galaxy.
A color index, $B-V$, is relatively small (0.45, NED).
UV band images, a sensitive probe for detection of young stellar clusters, show 
a clear spiral arm structure with bright regions (e.g., \cite{Bianchi2005,Kuntz2010}).
Most of the bright regions coincide with giant HII regions, sites of intense star formation qualified as starbursts. 

Previous X-ray observations of M101 with Einstein, ROSAT, XMM-Newton, and Chandra resolved many point sources and 
found diffuse X-ray emission with $kT<$1 keV
(\cite{Trinchieri1990,Wang1999,Pence2001,Kuntz2003,Tyler2004}; Jenkins et al. 2004, 2005; \cite{Warwick2007, Kuntz2010}).
The bulk of the diffuse soft X-ray emission is correlated with the UV emission \citep{Kuntz2003,Warwick2007,Kuntz2010}.
The spectral properties 
are similar to those of the giant HII regions \citep{Kuntz2003,Kuntz2010,Sun2012}.
These facts imply that the diffuse soft X-ray emission is associated with star-formation activity:
stellar winds from massive stars, SNe, and superbubbles \citep{Kuntz2003,Warwick2007,Kuntz2010}.

Due to the lack of the sensitivity in the hard X-ray band (Einstein and ROSAT) or 
the high detector background (Chandra and XMM-Newton),
it would be difficult to find a hotter component.
The X-ray Imaging Spectrometers (XIS) onboard the Suzaku satellite has better spectral resolution
and lower/more stable intrinsic background than 
the previous X-ray satellites \citep{Koyama2007a,Mitsuda2007}.
Especially, the XIS has the best sensitivity in the iron line band.
We analyzed Suzaku archival data of M101 and found an emission line from the highly ionized iron, 
a sign of a hot plasma with a temperature of several keV.
The results of a preliminary analysis were reported in \citet{Yamauchi2014}.
In this paper, we present the results of a further analysis and discuss the origin of the iron emission line. 
Throughout this paper, the quoted errors are at the 90\% confidence level.

%%
% Figures 1
%%
\begin{figure*}
  \begin{center}
        \includegraphics[width=16cm]{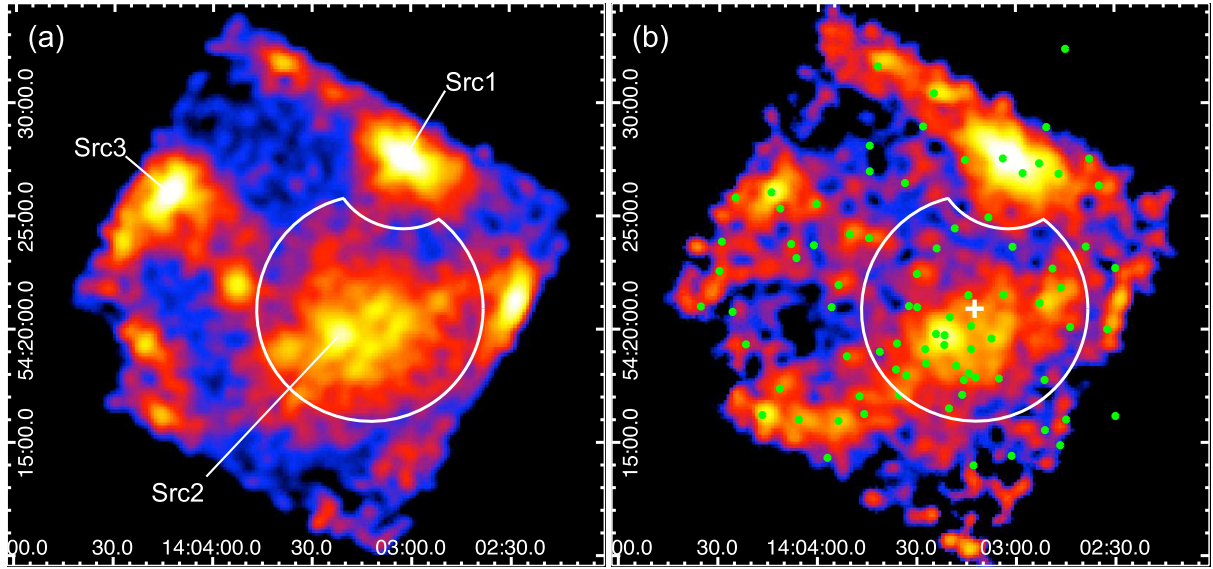}
  \end{center}
\caption{XIS image in the 0.5--2 (a) and 2--8 keV (b)  energy bands
smoothed with a Gaussian distribution.
The coordinates are J2000.0.  The data of XIS 0, 1, and 3 were co-added.
Non-X-ray background was subtracted and vignetting correction was performed.
The intensity levels are logarithmically spaced.
The green dots and the white cross in (b) 
show XMM-Newton sources (Jenkins et al. 2004, 2005) and the center of M101, respectively.
The white solid line shows a source region used in the spectral analysis.
 }\label{fig1}
\end{figure*}
%%%

\section{Observations and data reduction}

Suzaku observation of M101 was made with the XIS (\cite{Koyama2007a})
on the focal planes of the thin foil X-ray Telescopes (XRT, \cite{Serlemitsos2007})
on 2006 November 23--25 (Obs. ID=801063010).  
The Hard X-ray Detector (HXD: \cite{Takahashi2007, Kokubun2007})
observed M101 simultaneously, but only the XIS data were used because
we focus on an iron emission line from a thin thermal plasma.
The observed field of the XIS was a \timeform{17.'8}$\times$\timeform{17.'8} area with the center at 
($\alpha$, $\delta$)$_{\rm J2000.0}$= (\timeform{210.D8674}, \timeform{+54.D3599}).
The XIS consists of 4 sensors.
XIS sensor-1 (XIS 1) is a back-side illuminated CCD (BI), while
the other three XIS sensors (XIS 0, 2, and 3) 
are front-side illuminated CCDs (FIs).
One of the FIs  (XIS 2) stopped working on 2006 November 9.
Therefore, we utilized the XIS 0, XIS 1, and XIS 3 data for analyzing.
The XIS was operated in the normal clocking mode.

Data reduction and analysis were made using the HEAsoft version 6.13.
The XIS pulse-height data for each X-ray event were converted to 
Pulse Invariant (PI) channels using the {\tt xispi} software
and the calibration database version 2013-07-24.
We excluded the data obtained at the South
Atlantic Anomaly, during the earth occultation, and at the low elevation
angle from the earth rim of $<$ 5$^{\circ}$ (night earth) and $<20^{\circ}$
(day earth) and also removed hot and flickering pixels.
After the screening, we used the grade 0, 2, 3, 4, and 6 data.
The resultant exposure time was 98.9 ks.

\section{Analysis and results}

\subsection{Image}

Figure 1 shows X-ray images in the 0.5--2 and 2--8 keV energy bands.
For maximizing photon statistics, the data of XIS 0, 1, and 3 were added.
Two bright X-ray sources near to the northwest (Src1) and the northeast (Src3) edges were clearly found.
Their source positions were determined to be 
(RA, Dec)$_{\rm J2000.0}$=(\timeform{14h03m02s.6}, \timeform{+54D27'37''})
and
(RA, Dec)$_{\rm J2000.0}$=(\timeform{14h04m12s.8}, \timeform{+54D26'06''}), respectively.
The typical positional uncertainty of Suzaku is 19$''$ \citep{Uchiyama2008}.
In addition, the systematic error of the peak determination is 8$''$.
Taking the uncertainties into account, we identified these sources
with XMM-1 (XMMU J140303.9$+$542734)  and XMM-3  (XMMU J140414.1$+$542604) 
in Jenkins et al. (2004, 2005), respectively.
X-ray luminosities of XMM-1 and XMM-3 
have been estimated to be $>$10$^{39}$ erg s$^{-1}$ 
\citep{Jenkins2004}, and hence the two sources are possibly ultra luminous X-ray sources.  
We confirmed that both sources have high luminosities of $>$10$^{39}$ erg s$^{-1}$ in the present 
Suzaku observation (see Appendix).

A relatively bright X-ray source (Src2) was found near to the field of view (FOV) center.
The source position was determined to be 
(RA, Dec)$_{\rm J2000.0}$=(\timeform{14h03m21s.6}, \timeform{+54D19'49''}).
There are two possible counterparts within the error region, XMM-9 (XMMU J140321.6$+$541946) and 
XMM-14 (XMMU J140324.2$+$541949) (Jenkins et al. 2004, 2005). 
In addition to Src2, 
unresolved X-ray emission together with faint point sources is found in the central region of M101.

\subsection{4--10 keV band spectra}

X-ray spectra were extracted from a circular region with a radius of 5 arcmin, 9.9 kpc at the distance of 6.8 Mpc.
The contribution of the bright X-ray source, Src1, was excluded, 
but those of Src2 and the faint sources were included (see figure 1).
The non-X-ray background (NXB) was taken from the night earth data 
using {\tt xisnxbgen} \citep{Tawa2008}.
After subtracting the NXB, we merged the XIS 0 and XIS 3 spectra, but treated the XIS 1 spectrum 
separately because the response functions of the FIs and BI are different.
The XIS 0$+$3 and XIS 1 spectra were simultaneously fitted with the same model.
Response files, Redistribution Matrix Files (RMFs) and Ancillary Response Files (ARFs), 
were made using {\tt xisrmfgen} and {\tt xissimarfgen}, respectively.

Figure 2 shows the NXB-subtracted spectra in the 4--10 keV energy band.
At first, we fitted the spectra with a power-law (PL) function
and found positive residuals at 6--7 keV in both the FI and BI spectra.
We added an emission line (Line) model with a line width of null.
The $\Delta {\chi}^2$ value was 8.98, showing that 
the additional emission line model is statistically significant with a confidence level of 99 \%.
The best-fit parameters are listed in table 1 and the best-fit model is plotted in figure 2.
The center energy was determined to be 6.72$^{+0.10}_{-0.12}$ keV. 
The line is identified with a K-shell transition line from He-like iron, which 
indicates that the existence of an optically thin hot plasma with a temperature of several keV.
We derived an iron line luminosity, $L_{\rm Fe}$, to be 
$L_{\rm Fe}$=(8$\pm$4)$\times$10$^{37}$ erg s$^{-1}$ within the central 5$'$ radius region from the best-fit parameters.

%%
% Figures 2
%%
\begin{figure}[t]
  \begin{center}
      \includegraphics[width=8cm]{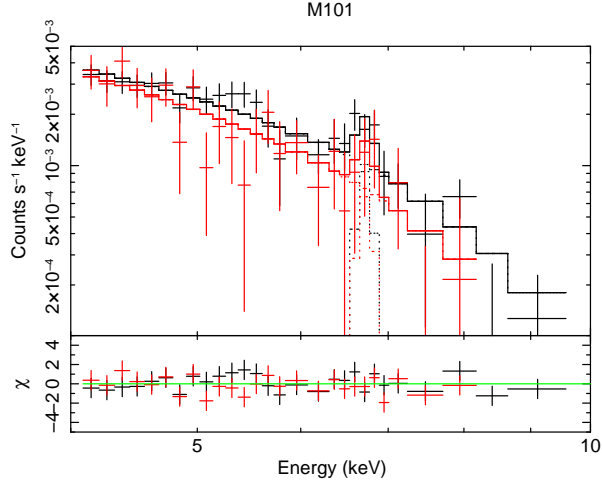}
        \end{center}
  \caption{
  XIS spectra in the 4--10 eV band within the 5$'$ radius region from the center of M101,
  black: XIS 0$+$3 and red: XIS1.
The histogram shows the best-fit power-law$+$emission line model (see table 1).
}\label{fig2}
\end{figure}

%%%%%%
% Table 1
%
\begin{table}[t]
\caption{The best-fit parameters for the 4--10 keV band spectra within the central 5$'$ radius region.}
\begin{center}
\begin{tabular}{lc} \hline  
Parameter & Value\\
\hline 
\multicolumn{2}{c}{Model: PL$+$Line}\\
\hline 
$\Gamma$   &  2.2$^{+0.4}_{-0.3}$  \\
$E_{\rm Line}$ (keV) & 6.72$^{+0.10}_{-0.12}$\\
$I_{\rm Line}^{\ast}$ (photons s$^{-1}$ cm$^{-2}$) & (1.3$\pm$0.7)$\times$10$^{-6}$ \\
$\chi^2$/d.o.f. & 36.8/50\\
\hline \\
\end{tabular}
\end{center}
\vspace{-10pt}
$^{\ast}$ Within the 5$'$ radius circle.\\
\end{table}
%%%%%%

\subsection{0.5--10 keV band spectra}

%%
% Figures 3
%%
\begin{figure*}[t]
  \begin{center}
      \includegraphics[width=8cm]{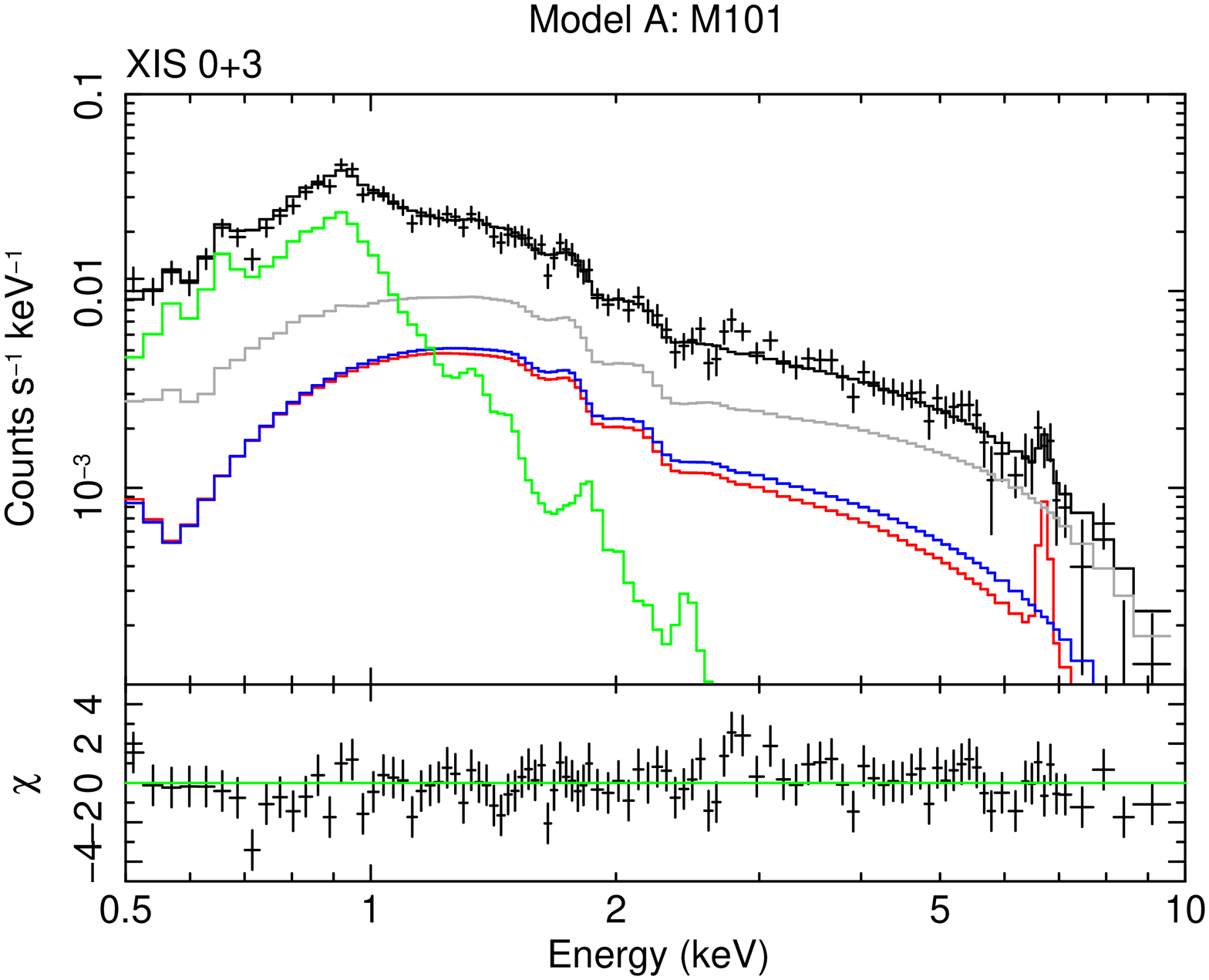}
      \includegraphics[width=8cm]{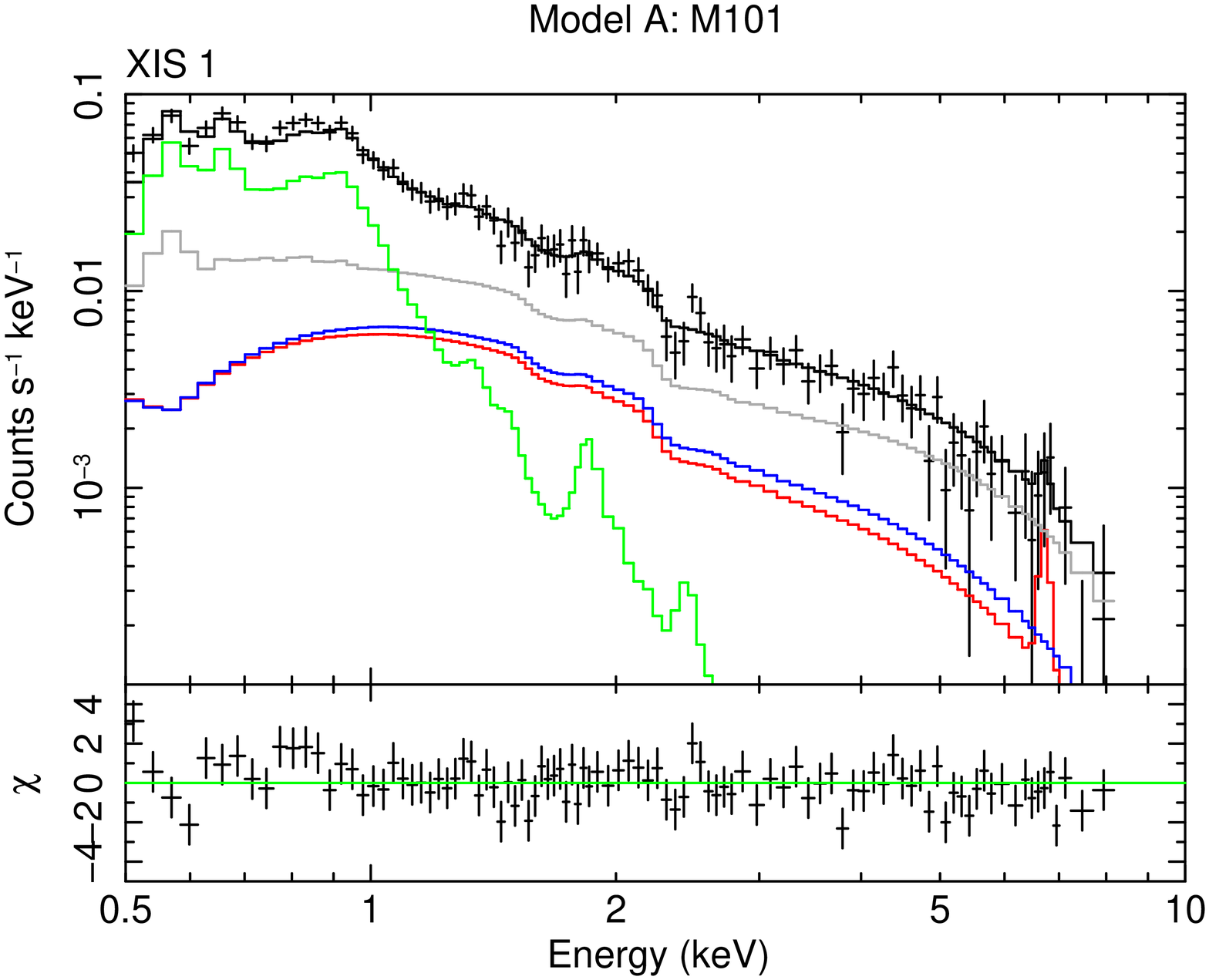}
      \includegraphics[width=8cm]{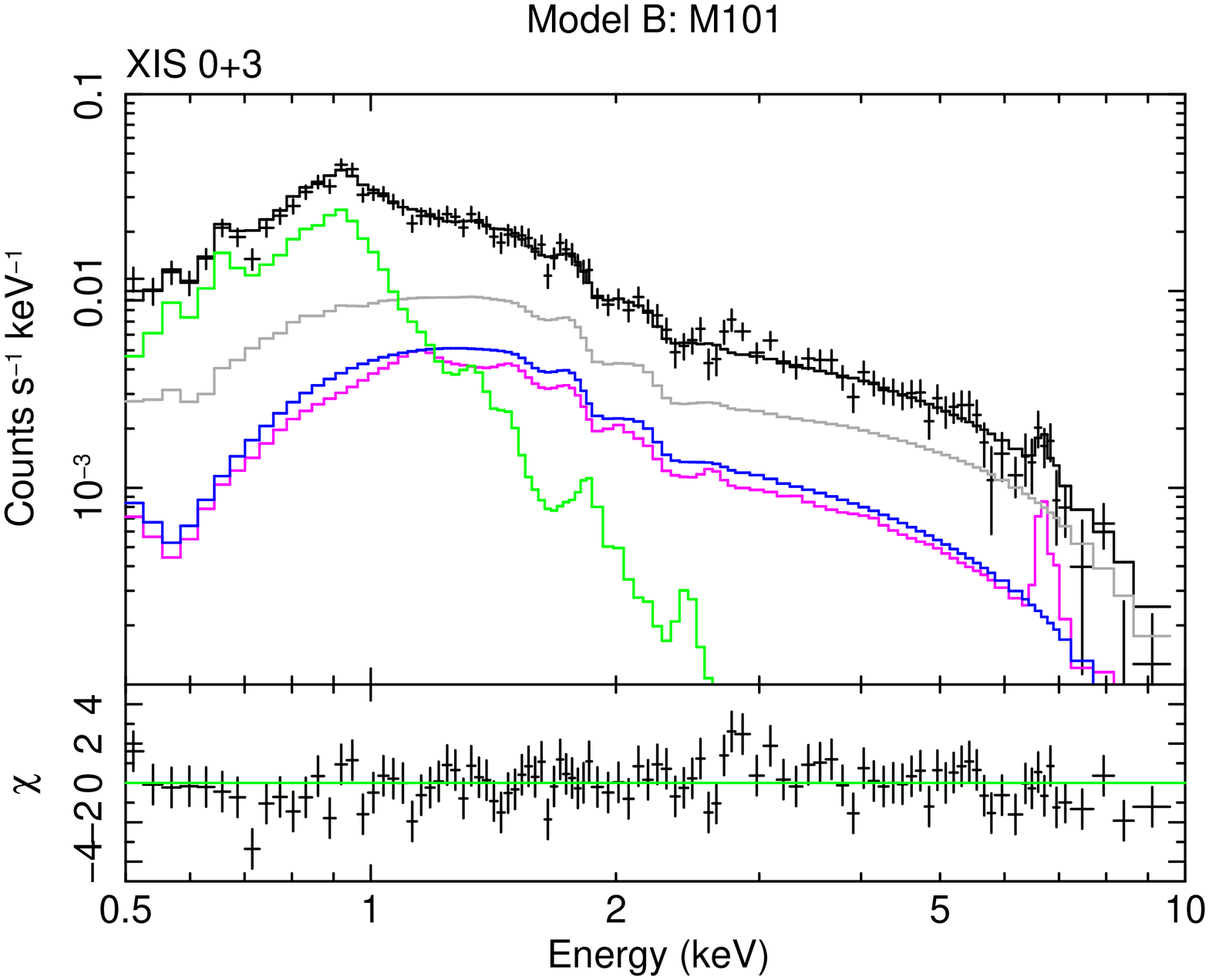}
      \includegraphics[width=8cm]{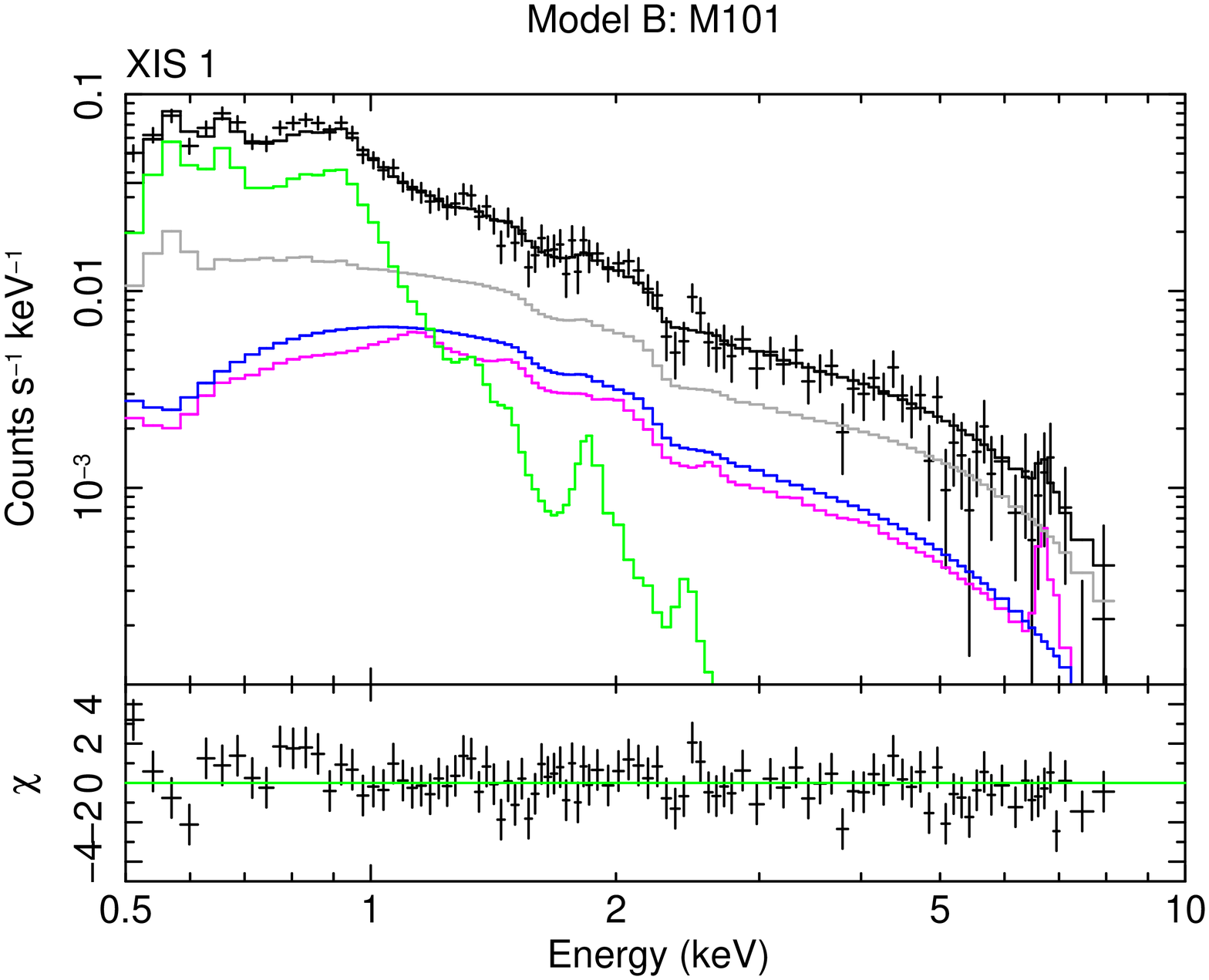}
  \end{center}
  \caption{XIS spectra in the 0.5--10 keV band within the  5$'$ radius region from the center of M101 (left: XIS 0$+$3 and right: XIS 1)
and the best-fit model (upper: Model A and lower: Model B).
Red, magenta, green, blue, and gray lines show the BR$+$Line, 
TP$_{\rm Fe}$, TP$_{\rm low}$$+$TP$_{\rm high}$, 
BR$_{\rm Src2}$, and Sky BGD models, respectively (see table 2).
}\label{fig3}
\end{figure*}

%%%%%%
% Table 2
%
\begin{table*}[t]
\caption{The best-fit parameters for the 0.5--10 keV band spectra within the central 5$'$ radius region.}
\begin{center}
\begin{tabular}{lcc} \hline  
Parameter & \multicolumn{2}{c}{Value}\\
\hline 
M101 & Model A & Model B\\
\hline 
$kT$ (keV)  & 5.1$^{+3.0}_{-1.5}$   & 6.9$^{+3.0}_{-1.5}$\\
Norm$^{\ast \dag}$ & (1.6$^{+0.3}_{-0.2}$)$\times$10$^{-4}$ &(1.2$\pm$0.1)$\times$10$^{-4}$ \\
$E_{\rm Line}$ (keV) & 6.72$^{+0.11}_{-0.13}$ & ---\\
$I_{\rm Line}^{\dag}$ (photons s$^{-1}$ cm$^{-2}$) & (1.0$^{+0.6}_{-0.7}$)$\times$10$^{-6}$ & ---\\
Abundance$^{\ddag}$ (Solar)& --- & 1 (fixed)\\
$kT_{\rm low}$ (keV)  &  0.187$^{+0.06}_{-0.08}$  & 0.187$^{+0.06}_{-0.08}$  \\
Abundance$_{\rm low}^{\ddag}$ & 1 (fixed) & 1 (fixed)\\
Norm$_{\rm low}^{\ast \dag}$ & (2.4$\pm$0.2)$\times$10$^{-4}$ & (2.4$\pm$0.2)$\times$10$^{-4}$ \\
$kT_{\rm high}$ (keV)  & 0.86$^{+0.03}_{-0.05}$ & 0.86$^{+0.03}_{-0.04}$  \\
Abundance$_{\rm high}^{\ddag}$ & 1 (fixed) & 1 (fixed)\\
Norm$_{\rm high}^{\ast \dag}$ & (4.0$\pm$0.5)$\times$10$^{-5}$ & (4.2$\pm$0.5)$\times$10$^{-5}$ \\
$kT_{\rm Src2}$ (keV) & 6.2 (fixed) & 6.2 (fixed)\\
Norm$_{\rm Src2}^{\ast}$ &  1.4$\times$10$^{-4}$ (fixed) & 1.4$\times$10$^{-4}$ (fixed)\\
$N_{\rm H}$  (cm$^{-2}$) &1.16$\times10^{20}$  (fixed) & 1.16$\times10^{20}$  (fixed)\\
\hline 
Sky BGD & &\\
\hline 
$kT_{\rm LHB}$ (keV) & 0.1 (fixed) & 0.1 (fixed) \\
Abundance$_{\rm LHB}^{\ddag}$ & 1 (fixed) & 1 (fixed)\\
Norm$_{\rm LHB}^{\ast \S}$ &  2.09$\times$10$^{-6}$ (fixed) & 2.09$\times$10$^{-6}$ (fixed)\\
$kT_{\rm MWH}$ (keV) &  0.27 (fixed)  & 0.27 (fixed)\\
Abundance$_{\rm MWH}^{\ddag}$ & 1 (fixed) & 1 (fixed)\\
Norm$_{\rm MWH}^{\ast \S}$ &   1.59$\times$10$^{-7}$ (fixed) &1.59$\times$10$^{-7}$ (fixed)\\
$\Gamma_{\rm CXB}$ &  1.412 (fixed)  & 1.412 (fixed)\\
Norm$_{\rm CXB}^{\P}$ & 8.17$\times$10$^{-7}$ (fixed) & 8.17$\times$10$^{-7}$ (fixed)\\
$N_{\rm H}$  (cm$^{-2}$) &1.16$\times10^{20}$  (fixed) & 1.16$\times10^{20}$  (fixed)\\
\hline
$\chi^2$/d.o.f. & 196.8/195 & 204.9/197 \\ \hline\\
\end{tabular}
\end{center}
\vspace{-10pt}
$^{\ast}$ Defined as 
10$^{-14}$$\times$$\int n_{\rm H} n_{\rm e} dV$ / (4$\pi D^2$),
where $n_{\rm H}$ is the hydrogen density (cm$^{-3}$), 
$n_{\rm e}$ is the electron density (cm$^{-3}$), and $D$ is the distance (cm).  \\
$^{\dag}$ Within the 5$'$ radius circle.\\
$^{\ddag}$ Relative to the solar value \citep{Anders1989}.\\
$^{\S}$ The unit is cm$^{-5}$ arcmin$^{-2}$.\\
$^{\P}$ The unit is photons s$^{-1}$ cm$^{-2}$ keV$^{-1}$ arcmin$^{-2}$ at 1 keV.\\
\end{table*}
%%%%%%

Next, we tried to fit the wide band spectra with a realistic model. 
Figure 3 shows the spectra in the 0.5--10 keV energy band.
The spectra contain the sky background and X-ray emission from M101.
Several authors have demonstrated that 
the sky background is typically represented by a three-component model,
a local hot bubble (LHB), a MW halo (MWH), and the Cosmic X-ray Background (CXB) 
(e.g., \cite{Konami2012,Akamatsu2013,Ota2013}).
According to the previous works, we used the following model, 
\begin{equation}
{\rm Sky\ BGD}={\rm TP_{LHB}} + ({\rm TP_{MWH}} + {\rm PL_{CXB}})\times{\rm ABS}, \\
\end{equation}
where TP is a thin thermal plasma emission ({\tt apec} model in XSPEC) and ABS is the photoelectric absorption by 
interstellar matter in the MW.
All the spectral parameters were fixed to values in \citet{Ota2013} (LHB and MWH) and \citet{Kushino2002} (CXB).

As described in section 3.2, we found a thermal component with a several keV temperature 
responsible for the iron line emission in M101. 
We modeled the emission as thermal bremsstrahlung (BR)$+$Line.
The source region contains the bright source Src2, whose
X-ray emission was represented by the BR model (see figure 4 and table 3).
The temperature and the normalization were fixed to values in table 3 (BR$_{\rm Src2}$).
The source region also contains other X-ray point sources with luminosities of less than $\sim$10$^{38}$ erg s$^{-1}$
(e.g., Jenkins et al. 2004, 2005; Warwick et al. 2007),
most of which would be neutron star low-mass X-ray binaries (NS-LMXBs).
The spectra of NS-LMXBs are well represented by a model consisting of a multicolor disk (MCD) model and a blackbody model 
\citep{Mitsuda1984,Makishima1989} and can be approximated by a BR model with a temperature of 
several keV to 10 keV.
Due to the spectral similarity,
it is difficult to distinguish X-ray emission of a thin thermal plasma with a several keV from that of NS-LMXBs in a spectral fitting.
Thus, the contributions of NS-LMXBs were included in the hottest component, the BR$+$Line model.
The Chandra and XMM-Newton observations showed the existence of the diffuse soft X-ray emission, composed of two or three 
thermal components
(e.g., \cite{Kuntz2003,Warwick2007,Kuntz2010}).
The Suzaku spectra also required at least two thin thermal components.
Thus, we modeled the diffuse soft X-ray emission as TP$_{\rm low}$ and TP$_{\rm high}$. 
We note that the model includes X-rays from soft X-ray sources such as stars, HII regions, 
and old supernova remnants (SNRs) in the source region.

The model for M101 (Model A) is expressed as follows, 
\begin{equation}
{\rm M101}=[({\rm BR}+{\rm Line})+{\rm TP}_{\rm low}+{\rm TP}_{\rm high}+{\rm BR_{\rm Src2}}]\times{\rm ABS}. \\
\end{equation}
The $N_{\rm H}$ value for ABS was fixed to 1.16$\times$10$^{20}$ cm$^{-2}$ \citep{Dickey1990}, while 
the cross sections of the photoelectric absorption were taken from Morrison and McCammon (1983). 
The abundances of the TP model were fixed to the solar values \citep{Anders1989}.
This model gave an acceptable fit (reduced $\chi^2$=196.8/195=1.009).
The best-fit model is plotted in figure 3, while the best-fit parameters are listed in table 2.
We derived $L_{\rm Fe}$ to be (6$\pm$4)$\times$10$^{37}$ erg s$^{-1}$  
from the best-fit parameters.

We also tried the spectral fitting by replacing BR$+$Line with another thin thermal plasma model, TP$_{\rm Fe}$ (Model B), 
\begin{equation}
{\rm M101}=({\rm TP}_{\rm Fe}+{\rm TP}_{\rm low}+{\rm TP}_{\rm high}+{\rm BR_{\rm Src2}})\times{\rm ABS}. \\
\end{equation}
The abundances were assumed to be solar values.
This model also gave an acceptable fit (reduced $\chi^2$=204.9/197=1.040).
The results are listed in table 2, while the best-fit model is plotted in figure 3.

\section{Discussion}

We found the 6.7 keV line from M101, which indicates the existence of 
a thin thermal plasma with a temperature of several keV.
Based on the results of the 4--10 keV and 0.5--10 keV band analyses, 
we obtain $L_{\rm Fe}$=(2--12)$\times$10$^{37}$ erg s$^{-1}$.
An equivalent width (EW) is estimated to be 1.0$\pm$0.7 keV from the BR$+$Line model fit.
Since the continuum emission includes X-rays from point sources such as NS-LMXBs,
the EW estimated above is a lower limit for the hot plasma. 
Here, we discuss the origin of the intense iron emission line.

\subsection{Point source}

X-ray binaries containing a black hole or a neutron star are bright in X-rays.
Their emission is generally optically thick or power-law-like, and the 6.7 keV line is absent or very weak  
(e.g., \cite{White1995,Tanaka1995,Asai2000}). Thus, the spectral properties are much different.
On the other hand, 
cataclysmic variables (CVs), active binaries (ABs), and young stellar objects (YSOs) in star forming regions 
are known to have a thin thermal emission with the iron emission line 
(e.g., \cite{Ezuka1999,Gudel1999,Yamauchi1996}).
Thus, the stellar sources with a luminosity of $<$10$^{34}$ erg s$^{-1}$ are thought to be potential candidates.

M101 is the spiral galaxy similar in morphological type to the MW.
From the similarity, we can expect that the properties of the X-ray source populations in the
two galaxies are similar. 
The cumulative luminosity density of the stellar sources
(CVs, ABs, and YSOs) in the MW is estimated to be
$L_{\rm 2-10 keV}$/$M_{\ast}$=(4.5$\pm$0.9)$\times$10$^{27}$ erg s$^{-1}$ $M_{\odot}^{-1}$ 
\citep{Sazonov2006}, 
where $L_{\rm 2-10 keV}$ is the luminosity in the 2--10 keV band and $M_{\ast}$ is the stellar mass.
Adopting the rotation velocity of 170 km s$^{-1}$ (see table 1 in \cite{Kuntz2003}, and references therein),
we can estimate the total mass within the central 5$'$ radius region to be 6.7$\times$10$^{10}$$M_{\odot}$.
Assuming all the mass is equal to the stellar mass,  
we obtain the total stellar source luminosity of $L_{\rm 2-10 keV}$=(2.4--3.6)$\times$10$^{38}$ erg s$^{-1}$.
Since M101 also contains interstellar matter and dark mater, 
this value is an upper limit of the stellar source luminosity.
The metal abundances and temperatures of ABs and YSOs observed in the MW are typically 
0.3 times solar and 2--5 keV, respectively (e.g., \cite{Gudel1999,Yamauchi1996}), 
while the spectra of CVs are represented by a multi-temperature plasma emission model ({\tt cevmkl} model in XSPEC)
with the maximum temperature of 20 keV and the metal abundance of at most 
$\sim$0.5 solar (e.g., \cite{Baskill2005,Ishida2009}).
Assuming the above spectral model,  
we can estimate $L_{\rm Fe}$ to be 3--4\% of $L_{\rm 2-10 keV}$. 
Then, we obtain $L_{\rm Fe}$=(0.7--1.4)$\times$10$^{37}$ erg s$^{-1}$,
which is lower than the observed value.
Thus, the well-known stellar sources cannot account for the observed $L_{\rm Fe}$.
 
\subsection{Diffuse hot plasma}

The intense 6.7 keV iron line is found in young and middle-aged SNRs.
The plasma with a temperature of 2--5 keV and the solar abundance \citep{Anders1989}
exhibits the intense 6.7 keV iron line with the EW of 1--2 keV.
Thus, hot plasmas produced by SN explosions are a possible candidate.
Active star formation in the past leads to multiple SN explosions, and then
a large amount of diffuse hot plasma would be produced.
Radio observations have found many HI holes and shells, probably produced by SNe or stellar winds (e.g., \cite{Allen1979}). 
One of them, near to giant HII region NGC 5462, is a large hole (size$\sim$1.5 kpc) surrounded 
by expanding HI shell with a kinetic energy of a few 10$^{52}$ erg \citep{Kamphuis1991}.
Furthermore, SNR MF83 and an X-ray source in NGC 5471B are suggested to be hypernova remnants  
\citep{Wang1999HN,Chen2002,Sun2012}.
These facts indicate huge energy release in the past.

Adopting the typical SNR luminosity of $L_{\rm 2-10keV}\sim$10$^{34-36}$ erg s$^{-1}$, 
(e.g., \cite{Tsunemi1986,Seward2000}),
we can estimate $L_{\rm Fe}$ of each SNR to be $\sim$7$\times$10$^{32}$--1$\times$10$^{35}$erg s$^{-1}$, and hence
the total number of SNRs, $N_{\rm SNR}$, to account for the observed $L_{\rm Fe}$ is 700--10$^5$.
Taking into account that a typical SNR age with a several keV temperature ($t$) is $\le$10$^{4}$ yr,
a SN rate is estimated to be $N_{\rm SNR}$/$t$$\sim$0.07--10 yr$^{-1}$.
This scenario requires a higher SN rate than the current value (0.02 yr$^{-1}$, Matonick \& Fesen 1997).

A magnetic field may play an important role. 
If the magnetic reconnection as seen on the solar surface is occurred in the galaxy scale, a large amount of 
hot plasma would be produced \citep{Tanuma1999}.
Furthermore, the magnetic field may confine the hot plasma \citep{Makishima1994}.
However, information on the magnetic field (the strength, location, and configuration) is too limited to discuss it in detail. 
Measurements of the magnetic field are very important for further investigation.

\subsection{Comparison with the GDXE}

The thin thermal emission with the intense iron emission line is a similar feature to the GDXE. 
Here, we compare our results with those of the GDXE.

A deep observation of the GDXE at ($l$, $b$)=(\timeform{0.D08}, \timeform{-1.D42}) has resolved 
almost the GDXE flux into point sources \citep{Revnivtsev2009} 
and the origin of the GDXE is proposed to be due to a superposition 
of many faint point sources such as CVs and ABs \citep{Revnivtsev2009,Yuasa2012}.
\citet{Yuasa2012} reported that the GDXE spectra were successfully fitted by a sum of the spectra of magnetic CVs and ABs.
The best-fit model requires a larger Fe abundance (0.6--0.9 solar) than observed values from CVs 
(e.g., \cite{Yuasa2010}) and ABs (e.g., \cite{Gudel1999}) in the solar vicinity.
This suggests that the well-known CVs and ABs cannot account for the iron line intensity of the GDXE, 
similar to the M101 results. 
CVs and ABs with higher metal abundances or 
other X-ray sources having more intense iron emission lines are required.

If the thin thermal emission mainly originates from the stellar sources, 
$L_{\rm Fe}$ would be proportional to the galaxy mass.
$L_{\rm Fe}$ of M101 is larger than that of the GDXE ($L_{\rm Fe}$$\sim$10$^{37}$ erg s$^{-1}$, 
the sum of the Galactic disk, the Galactic center, and the Galactic bulge components, 
\cite{Yamauchi1990,Yamauchi1993}), but 
M101 is slightly less massive than the MW (see table 1 in \cite{Kuntz2003}, and references therein).
This may suggest that a substantial amount of iron line emission originates from high energy phenomena except for stellar sources.

\section{Conclusion}

We found an intense 6.7 keV iron line from the face-on spiral galaxy M101, 
This indicates the existence of a thin thermal plasma with a temperature of several keV.
The iron line luminosity of the central region (5 arcmin radius) 
is estimated to be (2--12)$\times$10$^{37}$ erg s$^{-1}$, larger than that of the GDXE.
The line intensity is stronger than those attributed to stellar sources, while
the SNR scenario requires a higher SN rate than the current value. 
A large number of 
unresolved X-ray sources with an intense iron emission line may exist or unknown mechanisms may be at work.
In order to understand the origin of the hot plasma,
further observations with high spatial resolution and high sensitivity are encouraged.

\section*{Acknowledgement}

The author is grateful to all members of the Suzaku team. 
I thank Akihiko Tomita and Hiromitsu Takahashi for helpful discussion.
This research made use of the NASA/IPAC Extragalactic Database (NED) operated by
the Jet Propulsion Laboratory, California Institute of Technology, under contract with NASA. This work was supported by the Japan Society for the Promotion of Science (JSPS) KAKENHI Grant Number 24540232.

%%
% Figures 4
%%
\begin{figure}[t]
  \begin{center}
      \includegraphics[width=8cm]{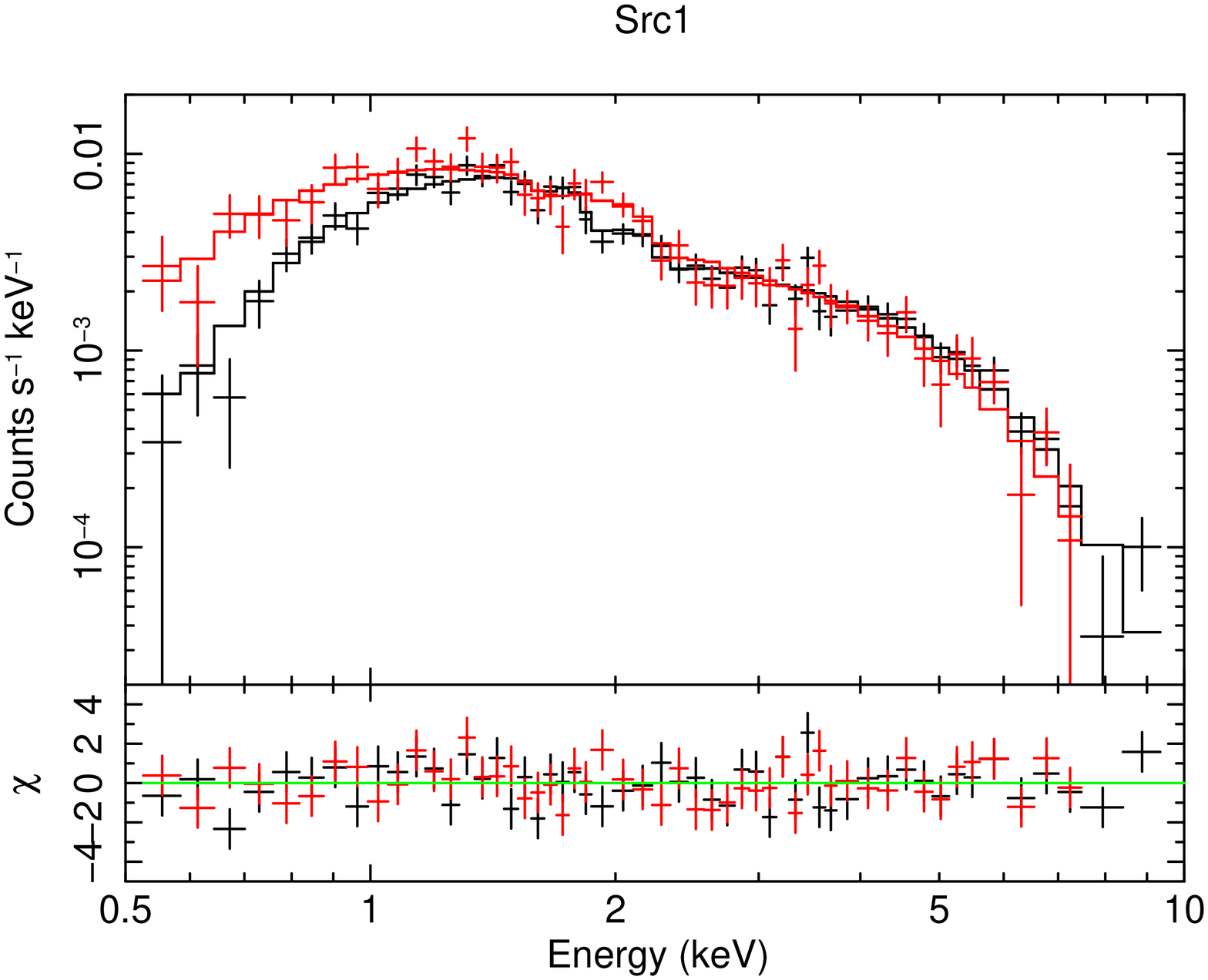}
      \includegraphics[width=8cm]{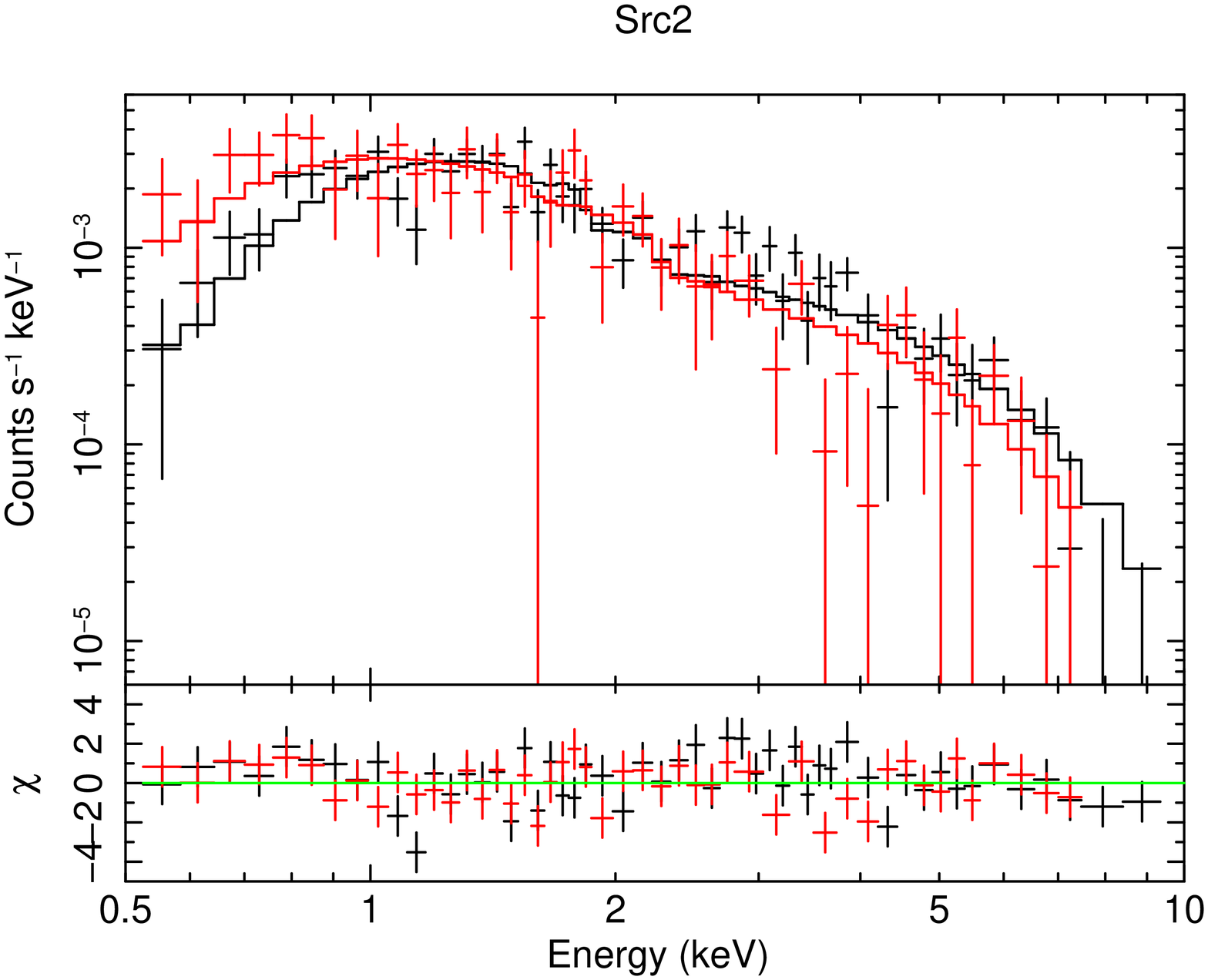}
      \includegraphics[width=8cm]{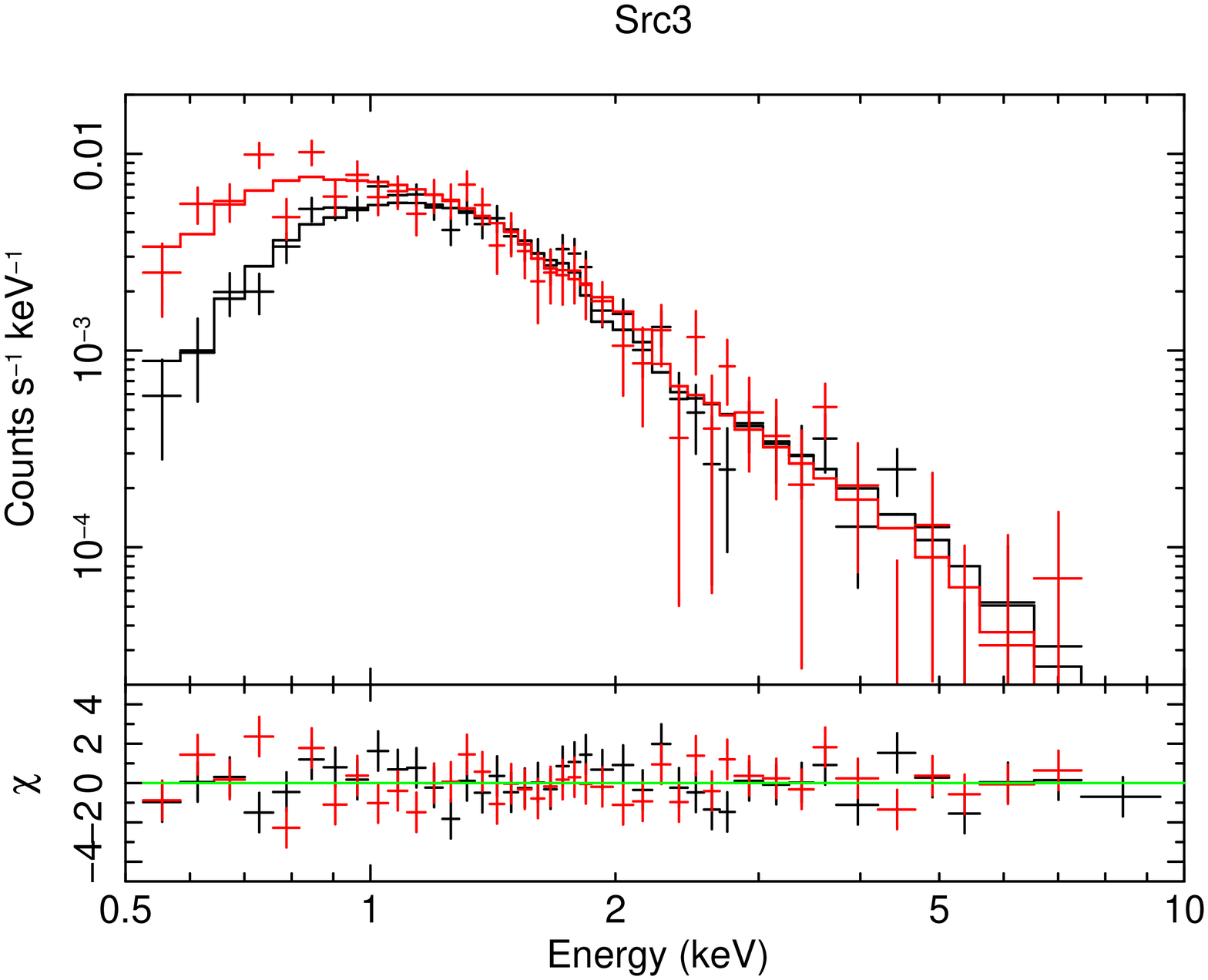}
  \end{center}
  \caption{XIS spectra of Src1 (top), Src2 (middle), and Src3 (bottom),
  black: XIS 0$+$3 and red: XIS1.
The histogram shows the best-fit model: multicolor disk, thermal bremsstrahlung, and 
power-law models for Src1, Src2, and Src3, respectively (see table 3).
}\label{fig4}
\end{figure}

\section*{Appendix: X-ray spectra of bright sources}

%%%%%%
% Table 3
%
\begin{table*}[t]
\caption{The best-fit parameters for 3 bright sources.}
\begin{center}
\begin{tabular}{lccc} \hline  
Source & Src1 & Src2 & Src3\\
 Source region & ellipse: 2$'$$\times$4$'$  & circle: 1$'$ radius & ellipse: 2$'$$\times$3$'$\\
 \hline
Parameter & \multicolumn{3}{c}{Value}\\
\hline 
\multicolumn{4}{c}{Model: MCD$\times$ABS}                     \\
\hline 
$N_{\rm H}$  ($\times10^{20}$ cm$^{-2}$) &7$\pm$3 & 1.16  (fixed)  & 1.16 (fixed)\\
$kT_{\rm in}^{\ast}$ (keV)  &  1.6$\pm$0.1  & 1.3$^{+0.2}_{-0.1}$ & 0.56$\pm$0.04\\
$r_{\rm in}$ $\sqrt{{\rm cos}\ i}$$^{\dag}$ & 47$^{+6}_{-5}$ & 36$\pm$6 & 235$^{+34}_{-30}$ \\
$\chi^2$/d.o.f. & 98.6/101 & 138.8/98 & 96.0/81\\
Observed flux$^{\ddag}$ ($\times10^{-13}$ erg s$^{-1}$ cm$^{-2}$)& 6.4 & 1.7 & 1.9\\
$N_{\rm H}$-corrected flux$^{\ddag}$ ($\times10^{-13}$ erg s$^{-1}$ cm$^{-2}$)&6.7 & 1.7 & 1.9 \\
 \hline
 \multicolumn{4}{c}{Model: BR$\times$ABS}                     \\
\hline 
$N_{\rm H}$  ($\times10^{20}$ cm$^{-2}$) &22$\pm$4 & 1.16  (fixed)  & 6$\pm$4 \\
$kT_{\rm e}$ (keV)  &  7.0$^{+1.3}_{-1.0}$ & 6.2$^{+1.7}_{-1.2}$ & 1.3$^{+0.3}_{-0.2}$\\
Norm$^{\S}$ & (5.6$\pm$0.3)$\times$10$^{-4}$ & (1.4$\pm$0.1)$\times$10$^{-4}$& (4.6$^{+1.0}_{-0.7}$)$\times$10$^{-4}$\\
$\chi^2$/d.o.f. & 101.1/101 & 126.8/98 & 79.0/80\\
Observed flux$^{\ddag}$ ($\times10^{-13}$ erg s$^{-1}$ cm$^{-2}$)& 6.8 & 1.9 & 2.0\\
$N_{\rm H}$-corrected flux$^{\ddag}$ ($\times10^{-13}$ erg s$^{-1}$ cm$^{-2}$)&8.0 & 2.0 &2.3 \\
 \hline
 \multicolumn{4}{c}{Model: PL$\times$ABS}                     \\
 \hline 
$N_{\rm H}$  ($\times10^{20}$ cm$^{-2}$) & 31$\pm$5 & 1.16  (fixed) &24$^{+6}_{-5}$\\
$\Gamma$ & 1.8$\pm$0.1  & 1.7$\pm$0.1 & 3.1$^{+0.3}_{-0.2}$\\
Norm$^{\P}$ & (1.6$^{+0.1}_{-0.2}$)$\times$10$^{-4}$ & (3.4$^{+0.3}_{-0.2}$)$\times$10$^{-5}$&(1.4$^{+0.2}_{-0.3}$)$\times$10$^{-4}$\\
$\chi^2$/d.o.f. & 117.5/101 & 132.8/98 & 74.4/80\\
Observed flux$^{\ddag}$ ($\times10^{-13}$ erg s$^{-1}$ cm$^{-2}$)&7.1 & 2.1 &2.1 \\
$N_{\rm H}$-corrected flux$^{\ddag}$ ($\times10^{-13}$ erg s$^{-1}$ cm$^{-2}$)& 9.0 & 2.1 &4.1 \\
\hline \\
\end{tabular}
\end{center}
\vspace{-10pt}
$^{\ast}$ Temperature at the inner-disk radius. \\
$^{\dag}$ $r_{\rm in}$ is the inner-disk radius at the 6.8 Mpc distance and $i$ is the inclination angle.\\
$^{\ddag}$ The 0.5--10 keV band flux calculated from the best-fit model. \\
$^{\S}$ Defined as 
10$^{-14}$$\times$$\int n_{\rm H} n_{\rm e} dV$ / (4$\pi D^2$),
where $n_{\rm H}$ is the hydrogen density (cm$^{-3}$), 
$n_{\rm e}$ is the electron density (cm$^{-3}$), and $D$ is the distance (cm).  \\
$^{\P}$ The unit is photons s$^{-1}$ cm$^{-2}$ keV$^{-1}$ at 1 keV.\\
\end{table*}
%%%%%%

We extracted the source spectra from the source region (see table 3), while
the background spectra was extracted from a nearby source free region in the FOV.
The NXB for the source and the background spectra were taken from the night earth data.
For both the NXB-subtracted source and background spectra, 
we made vignetting correction using the method described in \citet{Hyodo2008},
and subtracted the background spectra from the source spectra. 
The background-subtracted source spectra are shown in figure 4.
We tried to fit the spectra with 3 models: MCD, BR, and PL
models. 
When the $N_{\rm H}$ value was smaller than the Galactic absorption, 
the value was fixed to 1.16$\times$10$^{20}$ cm$^{-2}$.
The best-fit parameters are listed in table 3.
The spectral parameters and fluxes of Src1 (=XMMU J140303.9$+$542734) 
and Src3 (=XMMU J140414.1$+$542604) are roughly consistent with those in the XMM-Newton observations
\citep{Jenkins2004}.

%%%
% See the manual for the detail.
%%%

\end{document}